\documentstyle[12pt,epsf]{article}
 
\begin{document}
\begin{titlepage}

\setcounter{page}{1}
\rightline{WM-99-118}
\vfill
\begin{center}
 {\Large \bf Nonperturbative evaluation of the few-body states for 
scalar $\chi^2\phi$ interaction} 

\vfill
\vfill
 {\large \c{C}etin \c{S}avkl{\i}\footnote{csavkli@physics.wm.edu}
}

\vspace{.12in}
 {\it Department of Physics, College of William and Mary, 
Williamsburg, VA, 23187 }
\vspace{.075in}

\end{center}
\vfill
\begin{abstract}
A knowledge of nonpertubative propagators is often needed when the standard
perturbative methods are not applicable. An example of this is the bound 
state problem in field theory. While a nonperturbative result is valuable 
by itself, it is also an important guide for those who work on developing 
phenomenological models for the nonperturbative problem.  
The Feynman-Schwinger representation approach provides a convenient 
framework for calculating nonperturbative propagators. In this paper 
we provide an algorithm for computing 1,2, and 3 body bound states with the 
inclusion of all self energies, vertex corrections, ladder and crossed ladder 
exchanges. The calculation is done in the quenched approximation by ignoring 
the matter loops. We provide simulation results for 1,2 and 3-body states. 

\end{abstract}
\smallskip
\noindent
PACS codes: 12.38.-t\\
\noindent
Keywords: Nonperturbative, Monte-Carlo, bound states, Feynman-Schwinger 
representation

\end{titlepage}

\setcounter{footnote}{0}

\def\be{\begin{equation}}
\def\del{\partial}
\def\ee{\end{equation}}
\def\bea{\begin{eqnarray}}
\def\eea{\end{eqnarray}}
\def\ie{{\it i.e.}}
\def\eg{{\it e.g.}}
\def\half{{\textstyle{1\over 2}}}
\def\nicefrac#1#2{\hbox{${#1\over #2}$}}
\def\third{{\textstyle {1\over3}}}
\def\quarter{{\textstyle {1\over4}}}
\def\m{{\tt -}}

\def\p{{\tt +}}

\def\slash#1{#1\hskip-6pt/\hskip6pt}
\def\slk{\slash{k}}
\def\GeV{\,{\rm GeV}}
\def\TeV{\,{\rm TeV}}
\def\y{\,{\rm y}}

\def\l{\langle}
\def\r{\rangle}

\setcounter{footnote}{0}
\newcommand{\beqa}{\begin{eqnarray}}
\newcommand{\eeqa}{\end{eqnarray}}
\newcommand{\eps}{\epsilon}

\pagestyle{plain}
\setcounter{page}{1}
\section*{{Program Summary}}
{\em Title of the Programs:} phi3\\

\noindent
{\em Computer:} Sun 19\\

\noindent
{\em Operating system:} Unix\\

\noindent
{\em Programming language used:} FORTRAN 77\\

\noindent
{\em Peripherals used:} Laser Printer\\

\noindent
{\em Number of lines in distributed program:} 1510\\

\noindent
{\em Keywords:}  Nonperturbative, bound states, Feynman-Schwinger 
representation, Monte-Carlo, numerical evaluation.\\

\noindent
{\em Nature of physical problem:}\\
The program provided here evaluates the mass and distribution probabilities 
of the fully interacting n-body propagator ($n\le 3$) for scalar $\chi^2\phi$ 
interaction in 3+1 dimensions. The evaluation takes into account {\em all self 
energy, vertex dressing, and exchange interaction} contributions except those involving matter loops (the quenched approximation).
\\

\noindent
{\em Method of solution:}\\
The Feynman-Schwinger representation approach is used to express the 
field theoretical Green's function in terms of a quantum mechanical path 
integral. The resultant expression is evaluated using a Monte-Carlo simulation.
\\  

\noindent
{\em Restrictions of the program:}\\
Only $n=1,2$, and $3$ body propagators are considered. The extension to 
$n\ge 4$ requires straightforward modifications in the program.
\\

\noindent
{\em Typical running time:}\\
About 1 day for the 1-body propagator with the self energy.

\section*{{\bf LONG WRITE-UP}}
  
\section{Introduction}

In nuclear physics one is often faced by problems that require nonperturbative
 methods. The best known example is the problem of bound states. Even if the 
underlying theory may have a small coupling constant (such as in QED), and 
therefore allows the use of perturbation theory in general, the treatment of 
bound states are inherently nonperturbative. The n-body bound state is defined
by the pole of the interacting n-body propagator. A perturbative approximation
of n-body propagator does not produce the bound state pole location. 
Therefore it is essential that reliable nonperturbative methods that take all 
orders of interaction into account are developed. For this reason, numerous 
nonperturbative methods have been developed and successfully used in the 
literature. Some of the best known examples are lattice gauge 
theory~\cite{ROTHE} (LGT), and relativistic bound state 
equations~\cite{NAKANISHI,TJON0,GROSS1}. 

In a recent paper~\cite{SAVKLI1} we (along with authors J. Tjon, and F.
Gross) have discussed yet another method known as the Feynman-Schwinger 
representation(FSR). The basic idea in the FSR 
approach~\cite{SAVKLI1,SIMONOV1,TJON1,BRAMBILLA} is to integrate out all 
fields at the expense of introducing quantum mechanical path integrals over 
the {\em trajectories of particles}. Replacing 
the path integrals over fields with path integrals over trajectories has an 
enormous computational advantage. The advantage is due to the fact that 
the path integration over trajectories involves a variation of {\em lines}
rather than {\em fields in a volume}. Therefore the degrees of freedom is
considerably fewer. The FSR approach differs from the LGT in that it utilizes 
a space-time continuum, therefore maintaining the Poincare symmetry.    
On the other hand it should be pointed out that the FSR approach is not
without its drawbacks. In particular, how to extend the FSR approach to 
include fermions is not known. In the past 
researchers~\cite{BRAMBILLA,SIMONOV2,SIMONOV3} have attacked the fermion 
problem using various approximations. An {\em exact } result involving 
fermions is an important problem and requires further study.

While being able to calculate a nonperturbative result by itself is 
interesting, an additional motivation in studying the FSR approach is to 
determine which subsets of diagrams give the dominant contribution to the 
n-body propagator. This is particularly important in determining what kind of 
approximations are reasonable within the context of bound state equations. An 
example of how the FSR results can be used to compare different 
nonperturbative approximation schemes was presented in 
Refs~\cite{SAVKLI1,SAVKLI2}. In those works the emphasis was on the 
development of the formalism and application to the 1 and 2-body propagators. 
In Ref~\cite{SAVKLI2}, The FSR prediction for the
1-body mass in Scalar QED was compared by the rainbow-Dyson-Schwinger
equation prediction. It was found~\cite{SAVKLI2} that while the FSR approach 
provides a real mass pole for all coupling strengths, the Dyson-Schwinger 
equation provides a complex mass pole beyond a critical coupling strength. 
Furthermore it was found that, for Scalar QED in 0+1 dimension,  the vertex
corrections to the exchange interaction do not contribute to the 2-body 
binding energy~\cite{SAVKLI2}. These examples demonstrate the potential 
usefulness of the FSR approach. The knowledge about the nonperturbative 
propagators and vertices provided by the FSR is valuable as an input in 
testing and improving the modeling of other nonperturbative approaches such 
as Dyson-Schwinger equations.~\cite{ROBERTS1,SAVKLI0}  

The two and three-body bound state sectors provide possibilities for the
application of the FSR formalism. In particular it is 
important to see how various bound state equations (such as Bethe-Salpeter, Gross (spectator), Blankenbekler-Sugar, Equal-time, and nonrelativistic) 
compare with the quenched FSR results. Applications of the FSR approach to 2 
and 3 body states with comparisons to various bound state equation results in 
$\chi^2\phi$ theory is currently under study and will be
presented in a separate article.~\cite{SAVKLI3}

In this paper we present a complete numerical algorithm for the evaluation
of $n$-body masses ($ n\le 3 $) and distribution probabilities for scalar 
$\chi^2\phi$ interaction in 3+1 dimensions. By providing this algorithm
we intend to facilitate the comparison of various nonperturbative
methods with the exact quenched results in $\chi^2\phi$ theory. 
The organization of the paper is as follows: In the next section we present a 
brief summary of the FSR formalism. In the third section, using various
1, 2, and 3-body cases as examples, we discuss how results are obtained. And 
in the fourth section we explain the components of the program.    

\section{The Feynman-Schwinger representation for scalar fields}

\begin{figure}
\begin{center}
\mbox{
   \epsfxsize=3.0in
\epsffile{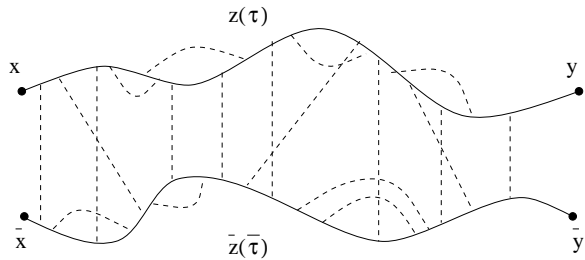}
}
\end{center}
\caption{A sample trajectory of each particle along with various interactions 
are shown. Except the matter loops all self energy, exchange energy and vertex
dressings are included. }
\label{trajectory}
\end{figure}

We consider the theory of charged scalar particles $\chi$ of mass $m$ 
interacting through the exchange of a neutral scalar particle $\phi$ of mass 
$\mu$. The Euclidean Lagrangian for this theory is given by 
\be
{\cal L}_E=\chi^*\bigl[m^2-\del^2+g\phi\bigr]\chi+\frac{1}{2}\,\phi(\mu^2-\del^2)\phi.
\label{lagr0}
\ee
The two body Green's function for the transition from the initial state
$\Phi_{i}=\chi^*(x)\chi(\bar{x})$ to final state $\Phi_{f}=\chi^*(y)\chi(\bar{y})$ is given by
\be
G(y,\bar{y}|x,\bar{x})=N\int {\cal D}\chi^*\int {\cal D}\chi\int {\cal D}\phi
\,\Phi^*_f\Phi_i\,e^{-S_E}.
\label{g0}
\ee
The final result for the two-body propagator 
involves a quantum mechanical path integral that sums up contributions coming 
from all possible {\em trajectories} of {\em particles}
\begin{equation}
G=-\int_0^\infty ds \int_0^\infty d\bar{s} \int ({\cal D}z)_{xy}\int ({\cal D}\bar{z})_{\bar{x}\bar{y}}\,e^{-S[Z]},
\label{g1.phi3}
\end{equation}
where $S[Z]$ is given by
\be
S[Z]\equiv -iK[z,s]-iK[\bar{z},\bar{s}]+V_{0}[z,s]+2V_{12}[z,\bar{z},s,\bar{s}]+V_{0}[\bar{z},\bar{s}].
\label{action}
\ee
where
\bea
K[z,s]&=&(m^2+i\epsilon)s-\frac{1}{4s}\int_0^1 d\tau \,\frac{dz_\mu(\tau)}{d\tau}\frac{dz^\mu(\tau)}{d\tau},\label{k1}\\
V_{0}[z,s]&=&\frac{g^2}{2}s^2\int_0^1d\tau\int_0^1d\tau' \,\Delta(z(\tau)-z(\tau'),\mu),
\label{v0.phi3}\\
V_{12}[z,\bar{z},s,\bar{s}]&=&\frac{g^2}{2}s\bar{s}\int_0^1d\tau\int_0^1d\bar{\tau}\, \Delta(z(\tau)-\bar{z}(\bar{\tau}),\mu).\label{v12.phi3}
\eea
Here the $V_{0}[z,s]$ term represents the self energy contribution, while
the $V_{12}[z,\bar{z},s,\bar{s}]$ term represents the exchange interaction (Fig.~\ref{trajectory}).
The interaction kernel $\Delta(x)$ is defined by
\bea
\Delta(x,\mu)&=& \int \frac{d^4p}{(2\pi)^4}\frac{e^{ip\cdot x}}{p^2+\mu^2}=\frac{\mu}{4\pi^2|x|}K_1(\mu|x|).
\label{kernel.phi3}
\eea
While we present expressions for the 2-body case, generalization 
to an arbitrary n-body system is trivial.
The bound state spectrum can be determined from the spectral decomposition of the two body Green's function
\be
G(T)=\sum_{n=0}^{\infty}c_ne^{-m_nT},
\ee
where T is defined as the average time between the initial and final states
\be
T\equiv\frac{1}{2}(y_4+\bar{y}_4-x_4-\bar{x}_4).
\label{tdef}
\ee
In the limit of large $T$, the ground state mass is given by
\begin{equation}
m_0=\lim_{T\rightarrow\infty}-\frac{d}{dT} ln[G(T)]=\frac{\int {\cal D}Z S'[Z]e^{-S[Z]}}{\int {\cal D}Z e^{-S[Z]}},
\label{groundstate}
\end{equation}
While this result in principle is correct, the convergence to the asymptotic
mass is slow due to the continuum contribution. The spectrum of the particle
involves the mass pole and a cut beyond this pole representing the continuum
contribution. Assuming that the constituents of the bound state are restricted
to be at equal times in the initial and final states, the Green's function
can be written as a function of time $T$, total displacement ${\bf R}$, and 
relative coordinate ${\bf r}$, G(T,{\bf R},{\bf r}).\footnote{Dependence
on the initial relative final coordinate ${\bf r}_0$ is implicit.} 

In order to eliminate the contribution of continuum states, introduce the 
Fourier transform,
\be
\tilde{G}(T,{\bf P},{\bf p})=\int d^3{\bf R}\, d^3{\bf r}\, e^{i{\bf P}\cdot{\bf R}+i{\bf p}\cdot{\bf r}}G(T,{\bf R},{\bf r})
\ee 
where ${\bf P}$ is the CM momentum, and ${\bf p}$ is the relative momentum between particles. Setting both ${\bf P}={\bf p}=0$ one has
\be
\tilde{G}(T,0,0)=\int d^3{\bf R}\, d^3{\bf r} \, G(T,{\bf R},{\bf r}),
\ee 
which eliminates the contribution of the continuum and projects out the s-wave
state. While an integration over ${\bf r}$ is not necessary for the elimination
of the continuum contribution, it is useful in eliminating the 
contribution of states with nonzero orbital angular momentum. 

\begin{figure}[t,h,b]
\begin{center}
\mbox{
   \epsfxsize=5.0in
\epsfbox{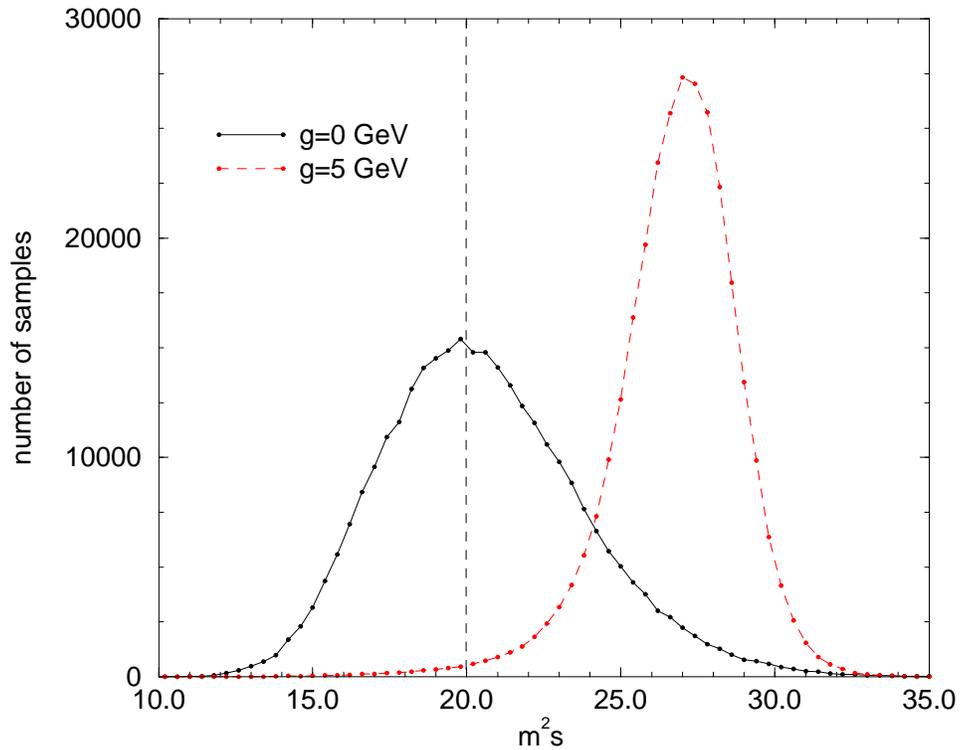}
}
\end{center}
\caption{The sample distribution of Feynman parameter s is shown for two different coupling strengths. The peak of the distribution moves from the 
classical value $T/2m$ as the coupling strength is increased.}
\label{sdist}
\end{figure}

\begin{figure}[t,h,b]
\begin{center}
\mbox{
   \epsfxsize=5.0in
\epsfbox{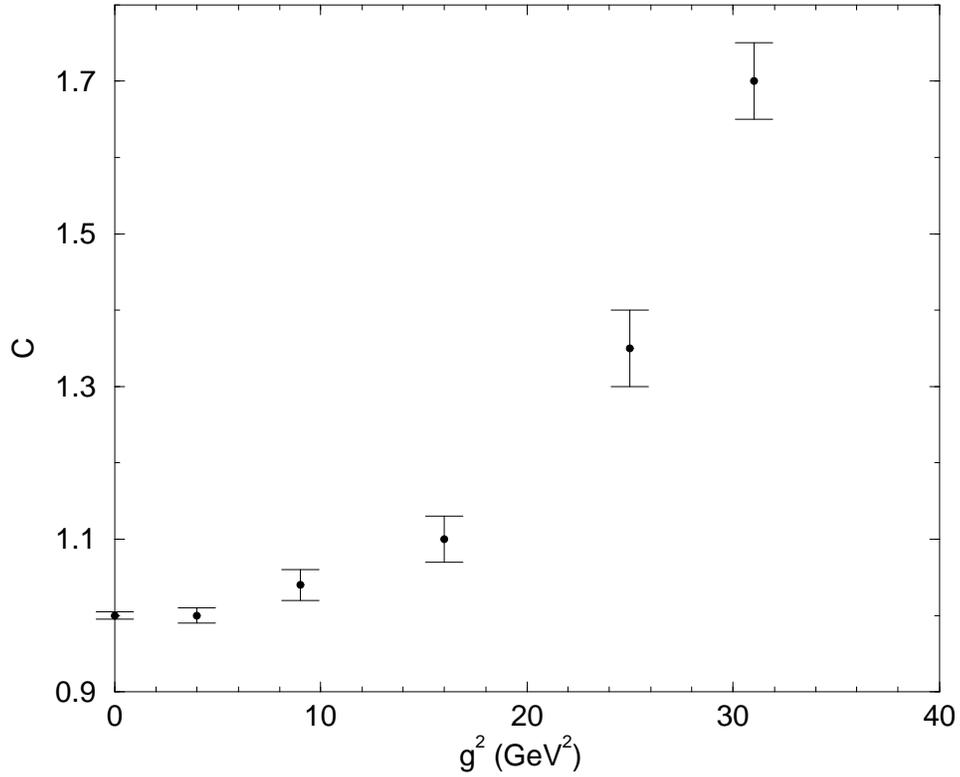}
}
\end{center}
\caption{The dependence of the peak of the s-distribution on the coupling 
strength is shown. The peak location is given by $s_0=C T/2m$. Beyond the 
critical coupling strenght of $g^2=31 {\rm GeV}^2$ a self consistent 
determination of $C$ is not possible. Therefore beyond the critical coupling strength 1-body mass becomes unstable.} 
\label{speak}
\end{figure}

While the result for the Green's function Eq.~\ref{g1.phi3} is exact in the 
quenched approximation, due to its oscillatory behavior it is not 
appropriate for Monte-Carlo simulation. In Ref.~\cite{SAVKLI1} it was shown 
that one can perform a Wick rotation in variable $s$ to avoid these 
oscillations. In the limit $g^2\rightarrow 0$ the dominant 
contribution to the integral in Eq.~(\ref{g1.phi3}) can be shown, by using the
saddle point method, to come from
\be
s=is_0=i\frac{T}{2m}.  
\ee
Since the large $s$ values do not contribute to the integral even without
 the interaction term, it is a good approximation to suppress the $g^2$ term 
at large s values. While this suppression is done it is important that 
the integrand is not modified in the region of dominant contribution 
$s\sim is_0$. This can be achieved by scaling the $s$ variable, {\em in the interaction term only}, by 
\begin{equation}
s\rightarrow \frac{s}{R(s,s_0)},
\label{sredef}
\end{equation} 
where
\begin{equation}
R(s,s_0)\equiv 1-(s-is_0)^2/\Gamma^2.
\label{rdef}
\end{equation}
In the free case, $(g^2=0)$, the width $W$ of the region of dominant $s$ 
contribution goes as 
\be
W=\sqrt{\frac{T}{2m^{3}}}.
\label{width}
\ee
Therefore, in the free case the dominant contribution to the $s$ integral 
comes from $i(s_0-W)<s<i(s_0+W)$. This claim is supported by the Monte-Carlo
simulation results. In Fig.~\ref{sdist} we present the results for 
$s$-distributions for two different coupling strengths for time $T=40$, and 
$m=1$ GeV. According to the estimate given above Eq.~(\ref{width}), for $g=0$, 
the dominant contribution to the  $s$-distribution comes from the region 
$15.53<s<24.47$ which is in agreement with the result presented in 
Fig.~\ref{sdist}. 
In order to ensure that the scaling given in Eq.~(\ref{sredef}) does not make 
a significant change in the region of dominant contribution, $\Gamma$ should 
be chosen such that
\be
\Gamma\ge W
\ee
As one increases the coupling strength, the value of $s_0$ deviates
from its free value. Therefore, in general, $s_0$ has to
be defined self consistently by monitoring the peak of the $s$ distribution.
In Figure~\ref{sdist} we display the s-distribution for two different coupling
strengths. It is seen that as the coupling strength is increased the peak of 
the s-distribution moves towards higher s values. In general the peak of the 
distribution can be parameterized as
\be
s_0=C\frac{T}{2m}.
\label{stationary}
\ee
The dependence of $C$ on coupling strength $g^2$ is determined self 
consistently. In Fig.~\ref{speak} $C$, which gives the location of the stationary point through Eq.~(\ref{stationary}), is plotted as a function of the 
coupling strength $g^2$. According to Fig.~\ref{speak}, it is not possible to 
find a self consistent stationary point beyond the critical 
coupling strength of $g^2=31$ GeV$^2$. A similar critical behavior was also 
observed in Refs.~\cite{ROSY} within the context of a variational approach.

The insensitivity of the dressed mass to the width $\Gamma$ has been 
investigated~\cite{SAVKLI1} and found that a choice of $\Gamma^2=2W^2$ was 
satisfactorily large. Results presented here employ the same value of $\Gamma$.

Prescription given by Eqs.~(\ref{sredef}), and (\ref{rdef}) enables one to 
perform a Wick rotation in variable $s$ to obtain a finite and non-oscillatory 
expression for the fully interacting two-body propagator:  

\begin{eqnarray}
G&=&\int_0^{\infty} ds \int_0^{\infty} d\bar{s} \int ({\cal D}z)_{xy}\int ({\cal D}\bar{z})_{\bar{x}\bar{y}}\label{g2}\\
&&\times{\rm exp}\biggl[-K[z,s]-K[\bar{z},\bar{s}]+V_0[z,s_r]+V_0[\bar{z},\bar{s}_r]+2V_{12}[z,\bar{z},s_r,\bar{s}_r]\biggr],\nonumber
\end{eqnarray}
where
\be
s_r\equiv \frac{s}{R(s,s_0)}.
\ee
The path integral is discretized using
\be
({\cal D})_{xy}\rightarrow(N/4\pi s)^{2N}\Pi^{N-1}_{i=1}\int d^4z_i,
\ee
where the s-dependence is critical in obtaining the correct normalization.
Note that the integration over the final coordinates is not included in this
expression.
 
The discretized versions of kinetic and interaction terms are given by
\bea
K[z,s]&\rightarrow& (m^2+i\epsilon)s-\frac{N}{4s}\sum_{i=1}^{N}(z_i-z_{i-1})^2,\\
V_0[z,s]&\rightarrow&\frac{g^2s^2}{2N^2}\sum_{i,j=1}^{N}\Delta(\frac{1}{2}(z_i+z_{i-1}-z_j-z_{j-1}),\mu),\\
V_{12}[z,\bar{z},s,\bar{s}]&\rightarrow&\frac{g^2s\bar{s}}{2N^2}\sum_{i,j=1}^{N}\Delta(\frac{1}{2}(z_i+z_{i-1}-\bar{z}_j-\bar{z}_{j-1}),\mu).
\eea

We next address the regularization of the ultraviolet (short distance) 
singularities. The ultraviolet singularity in the kernel $\Delta(x,\mu)$ 
Eq.~(\ref{kernel.phi3}) 
can be regularized using a Pauli-Villars regularization prescription. In order
to do this one replaces the kernel
\be
\Delta(x,\mu)\longrightarrow \Delta(x,\mu)-\Delta(x,\alpha\mu),
\ee
where $\alpha$ is in principle a large constant. The ultraviolet
singularity in the interaction is of the type
\be
\int dz\, z\, \Delta(z,\mu).
\ee
At short distances the kernel $\Delta(z,\mu)$ goes as $1/z^2$. Therefore,
the self energy calculation involves a logarithmic type singularity. The 
Pauli-Villars regularization takes care of this singularity. The 
Pauli-Villars regularization is particularly convenient for Monte-Carlo 
simulations since it only involves a modification of the kernel. In order to 
achieve an efficient convergence in numerical simulations we use a rather 
small cut-off parameter $\alpha=3$. This choice leads to a less singular 
kernel. The value of $\alpha$ can be increased arbitrarily at the cost 
additional computational time. It should be noted that while we employ a 
Pauli-Villars regularization throughout this paper, in general, 
the bound state problem {\em without} self energies {\em does not} have any 
UV singularities. UV singularities are only associated with the self energies
of the particles. After this brief summary of the formalism, in the next 
section we present some of the results and discuss the details of the 
algorithm.

\section{Running the code: applications}

\begin{figure}[t,h,b]
\begin{center}
\mbox{
   \epsfxsize=5.0in
\epsfbox{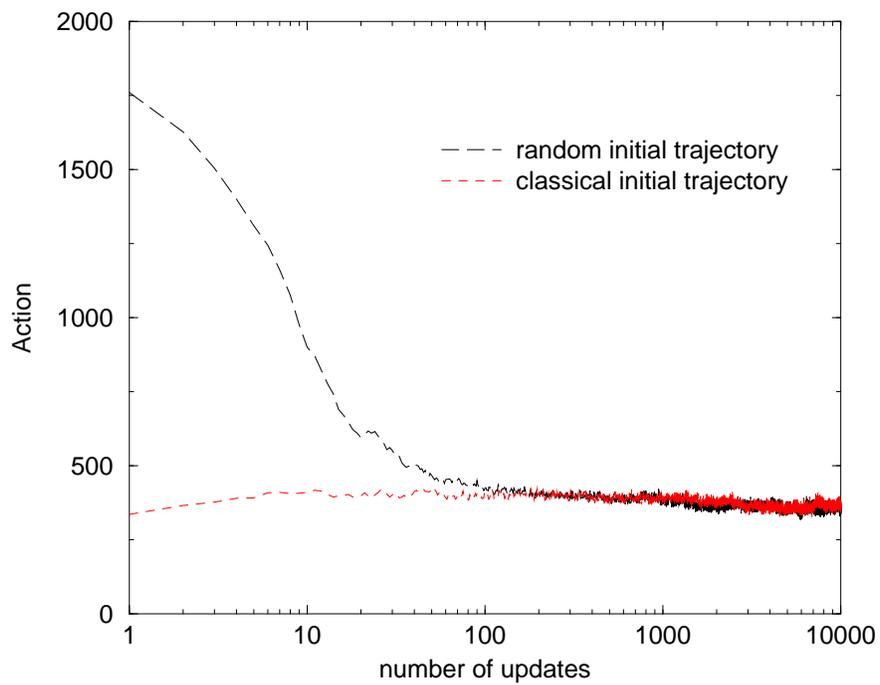}
}
\end{center}
\caption{The termalization for two different initial configurations is shown.
Irrespective of the initial conditions the action approaches to an asymptotic
limit. The first 1000 updates are ignored.   }
\label{termalize}
\end{figure}

\begin{figure}[t,h,b]
\begin{center}
\mbox{
   \epsfxsize=5.0in
\epsfbox{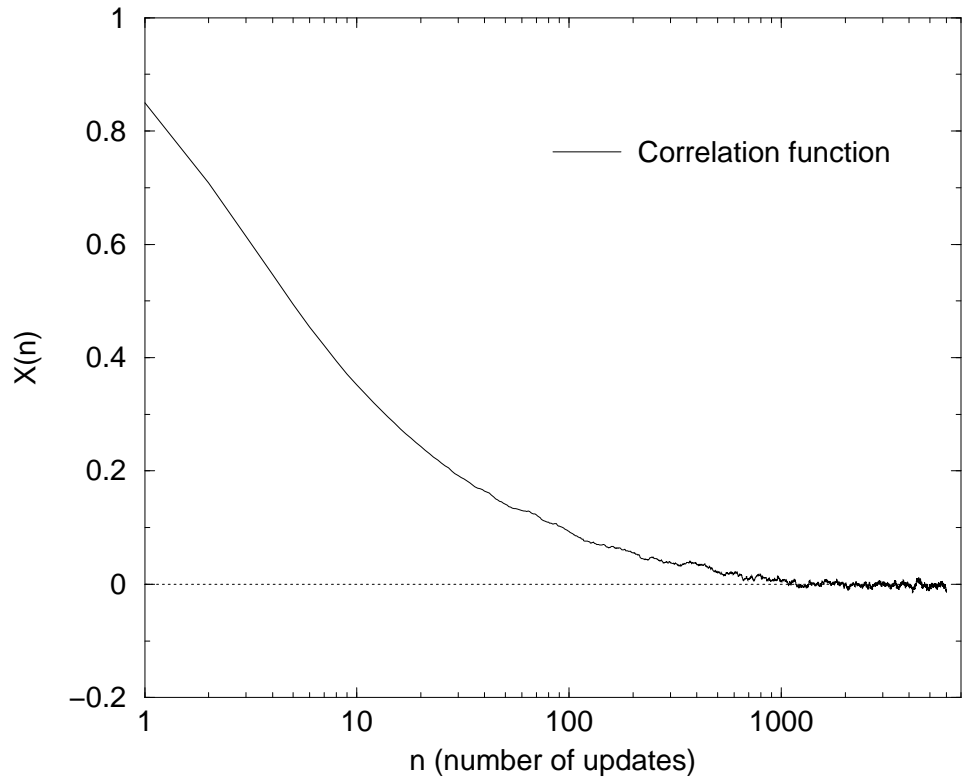}
}
\end{center}
\caption{Here we present the correlation function for the 1-body simulation. 
The number of updates necessary for the correlation to vanish is around 
$n=1000$. The typical number of updates performed in our simulations, which is 
around 500000, is well above the correlation length, thereby insuring that 
statistically uncorrelated configurations are sampled.}
\label{correlation}
\end{figure}

The Monte-Carlo simulation starts by choosing an initial configuration
for the trajectories of the particles. The choice of the initial 
trajectory is arbitrary (except at the end points). In Fig.~\ref{termalize}
the evolution of the action as a function of number of updates
for two different initial conditions is displayed. Starting with a random 
initial trajectory is analogous to a high temperature system and the 
termalization 
(reaching to the ground state) takes a long time (about 1000 updates). However
if one starts by a classical free trajectory, the initial configuration is 
analogous to a frozen system without any fluctuations and the termalization
results in an increase of the action. However asymptotically both results
should converge as shown in Fig.~\ref{termalize}. We usually start with an 
orderly system and disregard the first 1000 updates. Step sizes of the random 
walker in configuration space and in s-space should be chosen such that 
the average acceptance rate of Monte-Carlo updates are about \% 50. In each 
run we typically make 500000 updates of trajectories. In order to reduce the 
statistical errors runs must be repeated (usually more then 10 times) using 
different random number seeds. The correlation function $X(n)$ of sampled 
configurations in each run is defined as
\be
X(n)\equiv\frac{\left< m(i) m(i+n) \right> - \left< m \right>^2 }{\left< m 
\right>^2 },
\ee
where $m(i)$ is the mass measurement at the $i$'th update. The correlation 
function $X(n)$ measures how the information about a given configuration is
lost as a function of the number of updates $n$. In Figure~\ref{correlation} 
we show the correlation function for the 1-body problem. According to 
Fig.~\ref{correlation} the number of updates necessary for the correlation 
between sampled trajectories to vanish is around $n=1000$. The number of 
configurations sampled in our simulations, which is around 500000, is well 
above the correlation length 1000, thereby insuring that uncorrelated 
trajectories are sampled during the Monte-Carlo integration.

\begin{figure}[thb]
\begin{center}
\mbox{
   \epsfxsize=5.0in
\epsfbox{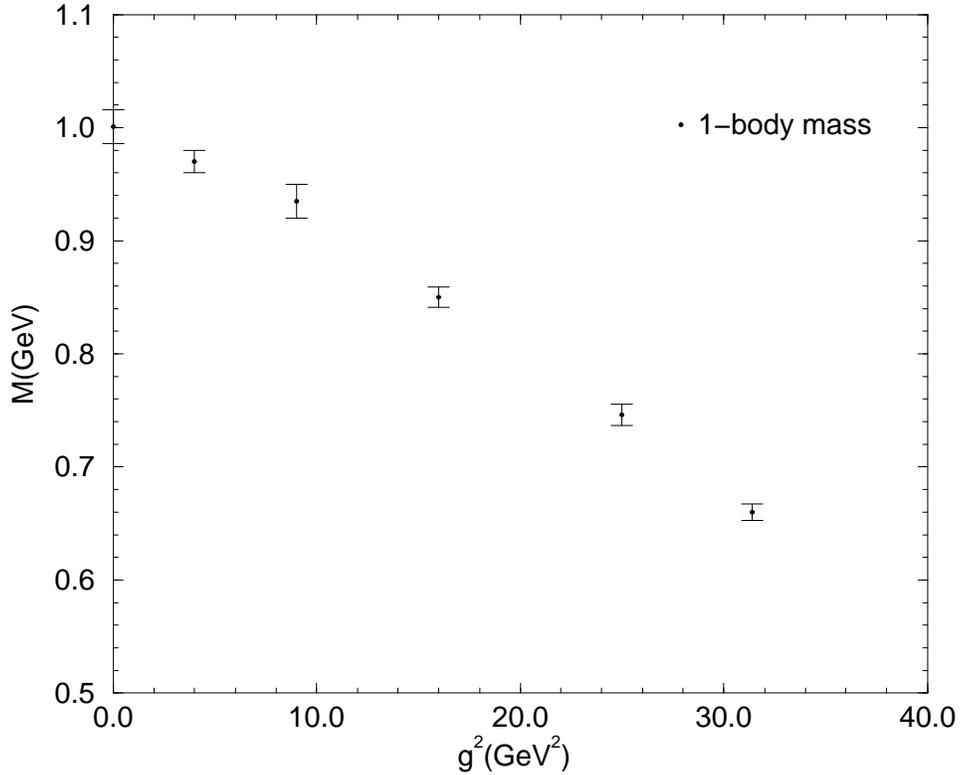}
}
\end{center}
\caption{The coupling constant dependence of the 1-body dressed mass is 
shown. The asymptotic time value $T$ used to obtain these mass values 
increase as $g$ increases. While at $g^2=0$ GeV$^2$ $mT=35$ is sufficient, at
 $g^2=31$ 
GeV$^2$ $mT=45$ is needed. The peak of the s distribution was self 
consistently determined. Beyond the critical coupling strength of $g^2=31$
 ${\rm GeV}^2$ the 1-body mass becomes unstable.}
\label{mvsg2.1b}
\end{figure}

The time $T$ required to reach to the asymptotic limit 
increases as  $g^2$ increases. For $g^2=25$ the asymptotic value of the mass, 
given by Eq.~(\ref{groundstate}), 
is obtained around $mT=40$. In particular for the result shown in 
Fig.~\ref{mvsg2.1b}, as $g^2$ ranged from $g^2=0$ GeV$^2$ to $g^2=31$ GeV$^2$ 
the asymptotic time values used were increased from $mT=35$ to $mT=45$. As 
time $T$ is increased the number of steps $N$ should also be proportionally 
increased so that the step size of the particle trajectory remains the same.  
As one increases the coupling strength $g^2$, trajectories of the particles 
deviate from the classical trajectory to a greater degree. This increase in 
fluctuations requires that one uses a higher $N$ value. The number of steps 
$N$ particles take between the initial and final coordinates is typically 
chosen to be $35<N<45$.

In Fig.~\ref{mvsg2.1b} the $g^2$ dependence of the one-body dressed 
mass is presented. 1-body masses presented here are lower than those in \cite{SAVKLI1}. This is due to the fact that in that work $C$ was assumed
to be approximately 1. However it was subsequently realized that $C$ deviates
from 1 significantly (see Fig.~\ref{speak}) as coupling strength is increased, 
and a self consistent determination of $C$ is important to pin down the 
critical point.  According to Fig.~\ref{speak} there is no self consistent 
stationary point Eq.~(\ref{stationary}) beyond the critical coupling strength 
$g^2=31$ GeV$^2$, and the 1-body dressed mass becomes unstable for 
$g^2>31$ GeV$^2$ (Fig.~\ref{mvsg2.1b}).

\begin{figure}[thb]
\begin{center}
\mbox{
   \epsfxsize=5.0in
\epsfbox{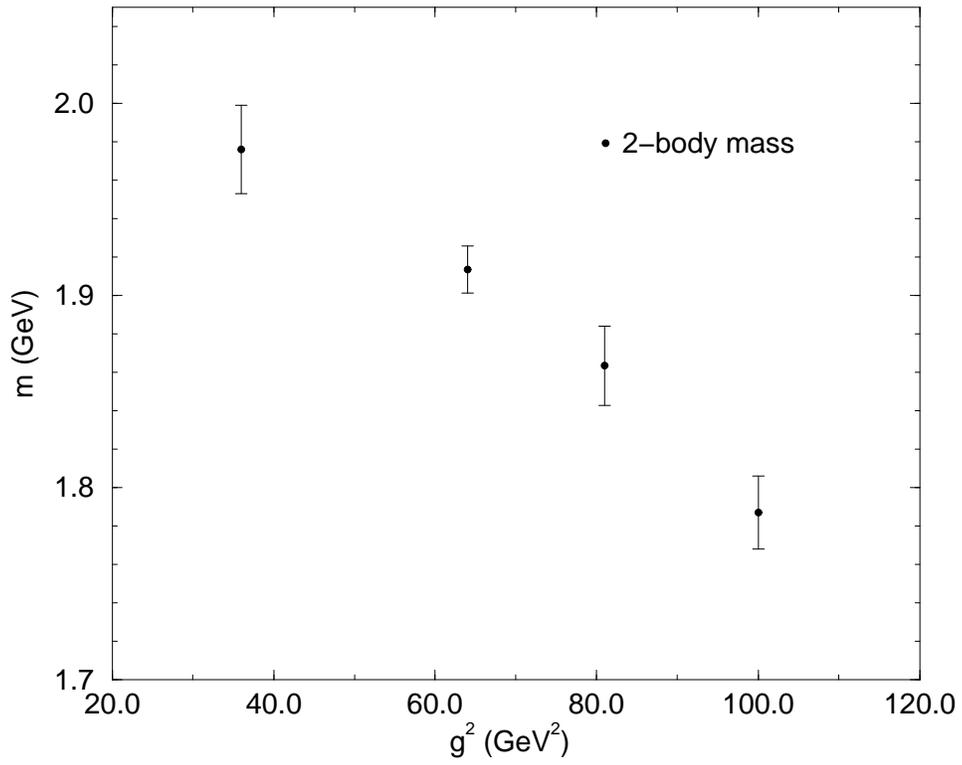}
}
\end{center}
\caption{The coupling constant dependence of the 2-body bound state mass is 
shown. Beyond the critical coupling strength of $g^2=100$ GeV$^2$ the 2-body
mass becomes unstable. }
\label{mvsg2.2b}
\end{figure}

\begin{figure}[t,h,b]
\begin{center}
\mbox{
   \epsfxsize=2.6in
\epsfbox{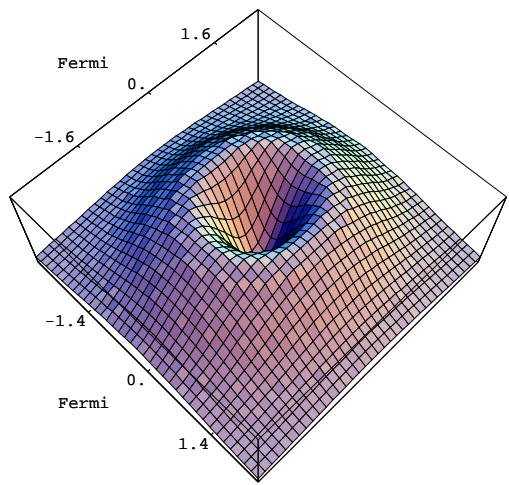}
}
\end{center}
\caption{The 2-body probability distribution when the 1st particles is at
the origin. The vanishing of probability at the origin is due to the phase 
space factor Eq.~(\ref{phase}).}
\label{rdist.surf.2b}
\end{figure}

\begin{figure}[t,h,b]
\begin{center}
\mbox{
   \epsfxsize=5.0in
\epsfbox{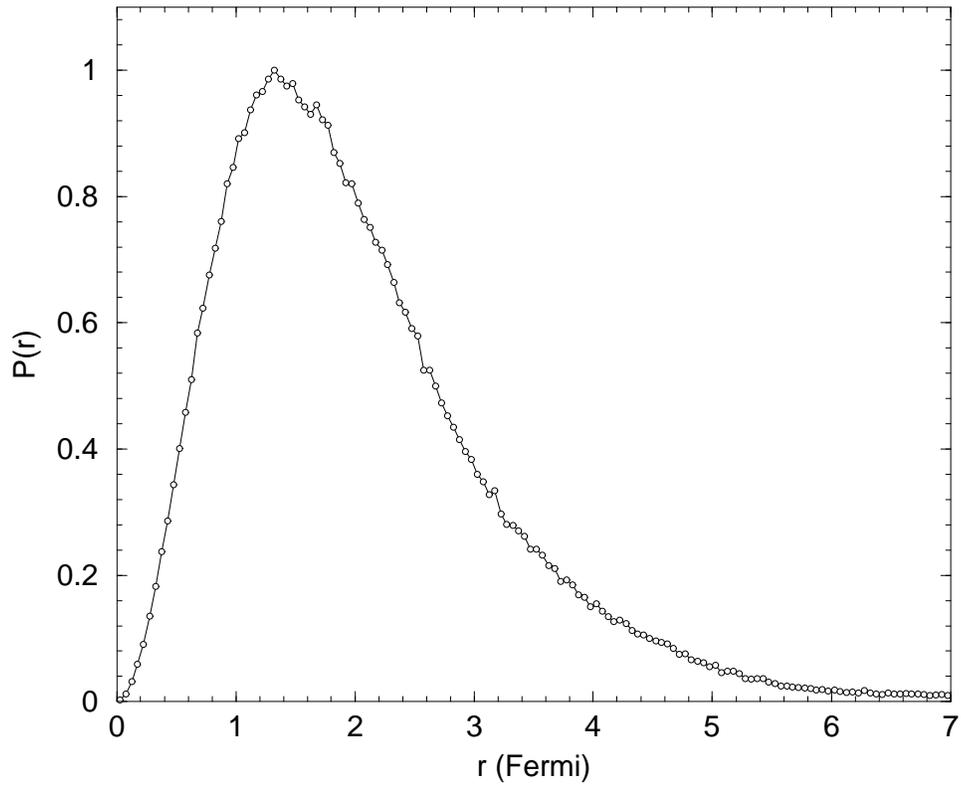}
}
\end{center}
\caption{The radial probability distribution for the 2-body bound state is 
shown. 
}
\label{rdist.2b}
\end{figure}

Next two applications involve the two and three-body bound 
states of equal mass particles. While the algorithm provided is capable to 
take into account all self energy and vertex dressing corrections to the bound
state, in the bound state applications provided here the self energy 
contributions of particles were not taken into account. This is controlled by 
a switch in the input file (see Table~\ref{PARAMET} for input options). In 
Fig.~\ref{mvsg2.2b} the $g^2$ dependence of the two-body bound state mass is 
presented. Beyond the critical coupling strength of $g^2=100$ GeV$^2$ the 
2-body mass becomes unstable. During the Monte-Carlo simulation the radial 
distributions of particles are stored in histograms. In general for an n-body 
bound state there are $n(n-1)/2$ relative distances. Since the particles are 
assumed to have equal time coordinates in the final state, essentially the 
relative distance is equal to the spatial distance. Let $r_{ij}=|r_i-r_j|$ be 
the distance between particles $i$ and $j$. In this case the histogram 
$P_{ij}(r_{ij})$ stores the number of final state configurations sampled in 
the interval $(r_{ij}-\delta/2,r_{ij}+\delta/2)$, where $\delta$ is the bin 
size. The range of $r_{ij}$ values is specified in the program using 
\be
0< r_{ij} < \frac{20}{m_i+m_j}\ \ {\rm Fermi}.
\ee
The choice of range is arbitrary as long as a reasonably smooth histogram is 
produced. The number of bins in each histogram is chosen to be 100. Therefore 
for two
particles of mass 1 GeV, the maximum radius to be histogrammed is 10 Fermi;
and the size of each bin is given by $\delta=0.1$ Fermi.  
In Fig.~\ref{rdist.surf.2b} we present the projection of the radial 
probability distribution onto a two-dimensional surface. The two-body 
probability distribution, presented in Fig.~\ref{rdist.surf.2b}, shows 
the result for $g^2=25$ GeV$^2$. Notice that the probability distribution 
vanishes at the origin. This is due to the phase space factor $4\pi r^2$. In 
order to be able to compare the probability distribution with a wave function 
the phase space factor should be factored out. In general the radial wave 
function {\em amplitude} $\Psi(r)$ is given by:
\be
|\Psi(r)|=\sqrt{\frac{P(r)}{4\pi r^2}}
\label{phase}
\ee
The normalization of the probability distribution histogram presented 
in Fig.~\ref{rdist.2b} is arbitrarily fixed such that the maximum entry is 
equal to 1. 

\begin{figure}[t,h,b]
\begin{center}
\mbox{
   \epsfxsize=5.5in
\epsfbox{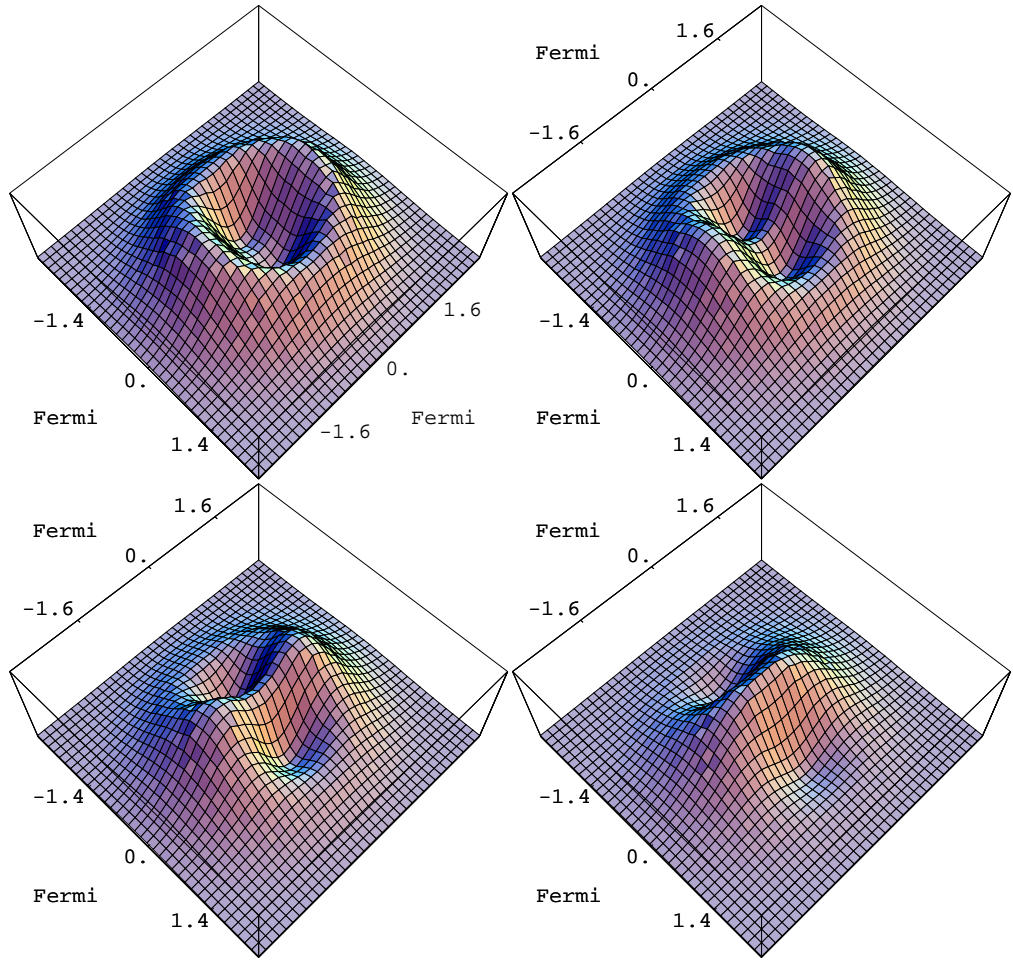}
}
\end{center}
\caption{Evolution of the probability distribution for the 3rd particle is
shown as the distance between the two fixed particles is increased. When the 
fixed particles are very close to each other the third particle sees them as a
point particle (the upper left plot). As the fixed particles are separated 
from each other the third particle starts penetrating between them (2nd and 3rd plots), and as the two fixed particles are maximally separated the third particle spends most of its time in between the two fixed particles (the lower 
right plot).
  }
\label{rdist.surf.3b}
\end{figure}

\begin{figure}[t,h,b]
\begin{center}
\mbox{
   \epsfxsize=2.0in
\epsfbox{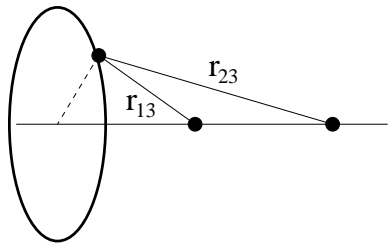}
}
\end{center}
\caption{The radial probability distribution for the 3-body bound state shown
in Fig.~\ref{rdist.surf.3b} includes the phase space factor represented by 
the ring in this figure. Distances $r_{13}$, and $r_{23}$ are fixed on the 
ring and each point on the ring contribute to the probability distribution shown in Fig.~\ref{rdist.surf.3b}.  }
\label{ring.3b}
\end{figure}

\begin{figure}[t,h,b]
\begin{center}
\mbox{
   \epsfxsize=5.0in
\epsfbox{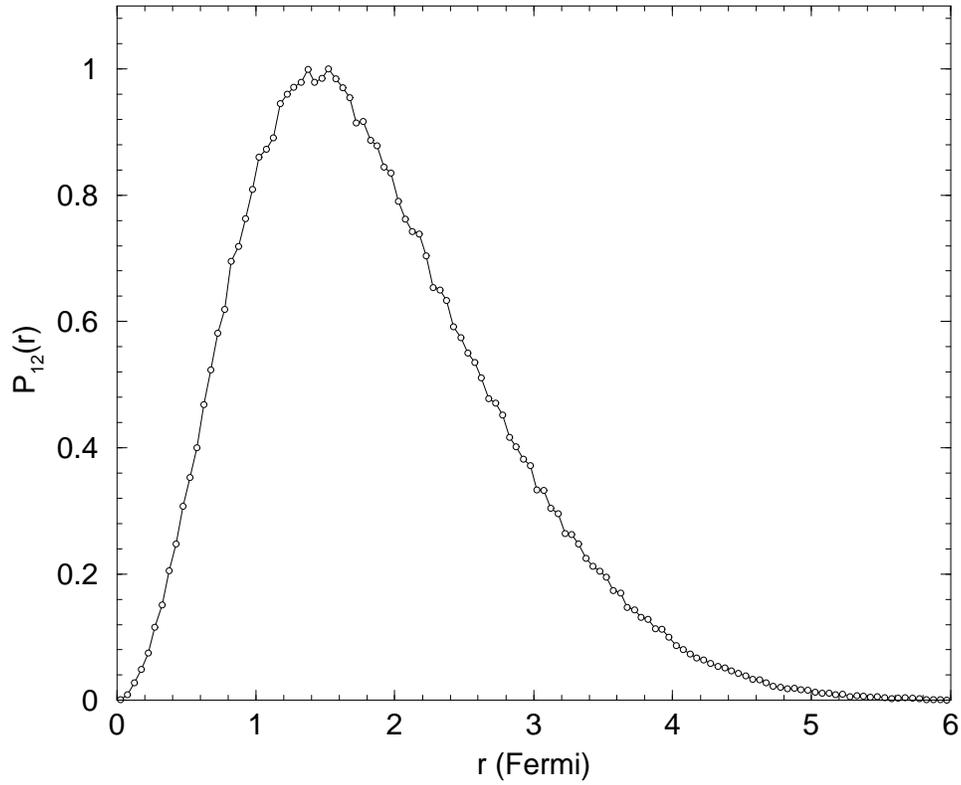}
}
\end{center}
\caption{The radial {\em two-body} probability distribution $P_{12}$ for the 
3-body bound state is shown. This distribution is obtained by integrating out 
the location of the 3rd particle.}
\label{rdist.3b}
\end{figure}

In generating the surface plot (Fig.~\ref{rdist.surf.2b}), it is assumed
that one of the particles is fixed at the origin. The amplitude 
of the surface gives the probability of finding the second particle at a 
distance $r$.
While the surface plot for a two-body bound state is not necessary for 
a visual understanding of the bound state structure, the three-body bound 
state demands a three dimensional plot. In the three-body case there are 
three relative coordinates. In order to be able to plot the three-body 
probability distribution we fix the location of two of the particles 
along the y axis, and calculate the probability of finding the third particle
in an arbitrary distance from two fixed particles. In Fig.~\ref{rdist.surf.3b}
we represent the probability distribution of the third particle for a given 
fixed configuration of the first and second particles. Assume that the fixed 
particles are particle 1 and particle 2. The probability distribution of 
the third particle is given by
\be
P_3(|{\bf r_3}|)\equiv P_{13}(|{\bf r_3-r_1}|)P_{23}(|{\bf r_3-r_2}|)
\ee  
Probability distribution $P_3(|{\bf r_3}|)$ includes the phase space contribution. Therefore $P_3(|{\bf r_3}|)$  represents
the probability of finding the third particle in a ring shown in 
Fig.~\ref{ring.3b}. For example in the first plot of Fig.~\ref{rdist.surf.3b}
two fixed particles are very close to each other such that the third particle 
sees them as a point particle. Therefore the probability distribution of
the first plot of Fig.~\ref{rdist.surf.3b} is very similar to the two-body
distribution given in  Fig.~\ref{rdist.surf.2b}. However as the fixed 
particles are separated from each other the third particle starts having 
a nonzero probability of being in between the two fixed particles (second
and third plots of Fig.~\ref{rdist.surf.3b}). Eventually when the 
two fixed particles are kept away from each other the third particle has 
a nonzero probability distribution only at the origin (the last plot 
shown in Fig.~\ref{rdist.surf.3b}). In Fig.~\ref{rdist.3b} the {\em two-body}
distribution function $P_{12}$ in a three-body system is shown.  All of these 
plots were produced with three equal particles of mass $1$ GeV, and the 
coupling strength of $g^2=64$ GeV$^2$.

\section{Program details}
In this section we discuss the details of the program. We start by providing
the tables of input parameters~\ref{PARAMET} and arrays~\ref{ARRAY} and then 
summarize the components of the program.

\subsection{Description of Input parameters}
\label{PARAMET}
\begin{flushleft}
\mbox{
\begin{tabular}{|l|l|} \hline
SIG = 1 GeV & Arbitrary momentum scale\\
\hline
IPAR(I)& I=1,2,3, Particles present  (0=n, 1=y)  \\
\hline
IEXCH(I,I)& I=1,2,3, Self  Interactions (0=n, 1=y)\\
\hline
IEXCH(I,J)& $(I,J)={(1,2),(1,3),(2,3)}$ Exch. Int. (0=n, 1=y) \\
\hline
QM(I) & Mass of particle I\\
\hline
G & The coupling strength\\
\hline
XMU & The exchange mass $\mu$\\
\hline
ALPHA & The Pauli-Villars mass ratio: $m_{pv}=\alpha$ $\mu$ \\
\hline
BETA = 1 & The width of $R(s,s_0)\equiv 1-\beta 
(s-iC \hat{s}_0)^2/T^\gamma$ \\
\hline
GAMMA = 1 & The T dependence of the width.\\
\hline
C & The peak location of the s distribution s=$C \hat{s}_0$.\\
  &  C is determined self-consistently. \\ 
\hline
N & Number of steps along the trajectory \\
\hline
NSMPL & $\#$ of sampled trajectories in MC integration \\
\hline
NVOID & No. of uncounted samples for initial termalization \\
\hline
ZSTEP & Max. step size of the random walker in coord. space \\
\hline
SSTEP & Max. step size of the random walker in $s$ space \\
\hline
EPSA  & The short distance(UV) cut-off for preventing\\
                        &   explicit occurrence of 1/0.0 type of singularity.\\
                        &  This is not a regulator however. The UV \\
                        &  regularization is done using Pauli-Villars subtraction\\
\hline 
IDUM & The seed of the random number generator \\
\hline
Z(I,0,J) & Initial coordinates of particle I=1,2,3   \\
\hline
Z(I,N,J) & Final coordinates of particle I=1,2,3  \\
\hline
IWRTA,IWRTM & Action, mass to be stored in file ? (0=n, 1=y)\\
\hline
INTOUT & Integrate over final coordinates ?\\
\hline
\end{tabular}
}
\end{flushleft}

\subsection{Arrays }\label{ARRAY}

\begin{flushleft}
\mbox{
\begin{tabular}{|l|l|} \hline
QM(3)  & Particle masses\\
\hline
IPAR(3)& Determines which particles are present. \\
\hline
   Z(3,0:400,4)& The trajectories of particles\\
\hline
ZNEW(3,0:400,4)& The updated trajectories of particles\\
\hline
SMAX(3) & The maximum value of s value to be histogrammed. \\
        &   (not an integration cut-off)\\
\hline
RMAX(3,3)& The maximum relative distance between particles\\
         & to be histogrammed. (not an integration cut-off) \\
\hline
SHIST(3,200)& Histogram of $s$ values \\
\hline
RHIST(3,3,200)& Histogram of relative distances between particles \\
\hline
SUMK(3) & Kinetic energies of particles\\
\hline
SUMV(3,3) & Potential energies of particles: SUMV(I,I): self\\
          & energy of the I'th particle, SUMV(I,J): the exchange \\
          & energy between I'th and J'th particles. \\
\hline
SUMKN(3)  & Updated kinetic energies\\
\hline
SUMVN(3,3) & Updated potential energies\\
\hline
\end{tabular}
}
\end{flushleft}

\subsection{Main Program}
The main program starts by calling the INPUT subroutine. This subroutine reads
input parameters described in Table~\ref{PARAMET}. 
Next, subroutine XINTGR is called to perform the Monte-Carlo simulation. In 
the following the role of each subroutine and function is explained. 

\subsection{Subroutine INPUT}
In this subroutine 12 input parameters are read. In addition to 
reading these parameters, histograms for the 
radial and s distributions, and initial trajectories of the particles are 
initialized. Initialization of trajectories is done using the classical 
free trajectories of particles. 

\subsection{Subroutine XINTGR }

 This subroutine {\em controls} basic steps of the Monte-Carlo sampling
process. First, the kinetic and the potential sums corresponding to the 
initial trajectories of particles are calculated by calling the XSUMS 
subroutine. Next step is the termalization process. In order to reach 
termalization the configuration of trajectories are updated NVOID times by 
calling the UPDATE subroutine. First NVOID updates are not used in the 
actual calculation of the bound state mass or the probability distribution. 
After the termalization the sampling is done for NSMPL times. The results
for the $s$ values and the relative distances of particles are 
histogrammed. Finally the bound state mass is calculated.
 
\subsection{Subroutine UPDATE }

This subroutine is responsible for updating the current configuration
of trajectories. Each coordinate and $s$ parameter is updated once.
Samplings are done according to distribution
\be
e^{-S[z]}
\ee
distribution. If the ratio $r$,
\be
r=e^{-(S[z']-S[z])},
\ee
is larger than 1 updates are always accepted. When $r<1$ the update is 
accepted with a probability of $r$.
\subsection{Subroutine XSUMS }
 
 This subroutine calculates kinetic and potential sums before the 
configuration updates start. Since calculation of the kinetic and potential 
sums is a costly operation, after every update we calculate the {\em shift} in the sums.

\subsection{Subroutine ACTION }

This subroutine calculates the action using known values of kinetic and
potential sums. The action is stored in a file to monitor the termalization. ( See Fig.~\ref{action} ).

\subsection{Subroutine UPDSUM }

Since calculation of the kinetic and potential sums is a costly operation,
after every update we calculate the {\em shift} in the sums. Using this 
shift sums are updated.
  
\subsection{Function XDERIV }

This subroutine calculates the derivative operator $S'[Z]$ 
Eq.~(\ref{groundstate}) which gives the mass of the bound state.

\subsection{Function DELTA, DELTAP }

These subroutine calculate, respectively, the interaction kernel 
Eq.~(\ref{kernel.phi3}) and its time derivative.

\subsection{Function DLARAN }

This is a random number generator obtained from LAPACK package at 
Netlib.~\cite{random} DLARAN returns a random real number from a 
uniform (0,1) distribution. In the actual implementation of the program
we have used a Numerical Recipes random number generator (ran2). However for 
copyright reasons here we provide this alternative random number generator. 

\subsection{Functions BESK0, BESK1, KZEONE }

BESK0 and BESK1 are modified bessel functions $K_0(x)$ and $K_1(x)$ which
are needed in calculation of the interaction kernel and its derivative.
BESKO and BESK1 call KZEONE subroutine~\cite{kzeone}. KZEONE subroutine 
returns real and imaginary parts of $e^xK_0(x)$ and $e^xK_1(x)$. The KZEONE
subroutine is considerably slower than the Numerical Recipes subroutines
for modified bessel functions. For copyright reasons we provide KZEONE rather
than the Numerical Recipes subroutines for the modified bessel functions. 

\section{Conclusions}

In this paper, using the Feynman-Schwinger representation, we have presented 
an algorithm that calculates 1,2,3 body masses and distribution probabilities 
in the quenched approximation for $\chi^2\phi$ theory in 3+1 dimension. The 
FSR approach provides an efficient method to calculate nonperturbative 
propagators. In this work we have presented results of applications to the 
1, 2 and 3 body states. A detailed comparison of quenched bound state results 
of the FSR approach with the bound state equation predictions is under study
and will be presented in a separate physics article~\cite{SAVKLI3}. It is hoped that, 
through comparison, this simple and rigorous nonperturbative method will 
enhance our understanding of various nonperturbative bound state models and 
approximations.  

\centerline{\bf Acknowledgements}
We are grateful to F. Gross, and J. Tjon for discussions. The support 
of the DOE through grant No. DE-FG02-97ER41032 is gratefully acknowledged.
The Thomas Jefferson National Accelerator Facility is gratefully acknowledged 
for warm hospitality and for providing computer resources.

\section{Test run INPUT}

\begin{tabular}{|l|l|}\hline
1.0 &  SIG     (GeV)  \\          
1 0 0 &  IPAR(1), (2), (3)\\
1 0 0 &  IEXCH(1,1)(2,2)(3,3)\\
0 0 0 &  IEXCH(1,2)(1,3)(2,3)\\
1.0 &  QM(1)  \\
1.0 &  QM(2)  \\
1.0 &  QM(3)  \\
2.0 &  G      \\
0.15&  XMU    \\
3.0 &  ALPHA  \\
0.5 &  BETA   \\
1.0 &  GAMMA  \\
1.0 &  C      \\
35  &  N      \\
500000&    NSMPL\\
5000  &  NVOID  \\
2.1 & ZSTEP   \\ 
4.5 & SSTEP    \\
1.0D-4&  EPSA   \\
1     &  IDUM             \\
 0. 0. 0. 0.& Initial coordinates\\
 1. 1. 0. 0.& ``\\
-1. 1. 0. 0.& ``\\\
 0. 0. 0. 35.& Final coordinates\\
 1. 1. 0. 35.& ``\\
-1. 1. 0. 35.& ``\\
0 0   &  IWRTA, IWRTM  \\
1     &  INTOUT \\
\hline
\end{tabular}

\section{OUTPUT}

\begin{tabular}{|l|l|}\hline
 SIG &    1.0 GeV\\
  
 Particles present & 1  0  0\\
  
 Self interactions &  1  0  0\\
  
 Exchange  "   1-2 &  0\\
    "      "   1-3 &  0\\
    "      "   2-3 &  0\\
  
 QM(1) &  1. GeV    \\
 QM(2) &  1. GeV   \\ 
 QM(3) &  1. GeV   \\ 
  
 G     & 2. GeV   \\ 
 XMU   &  0.15 GeV   \\ 
  
 ALPHA &  3.\\
  
 BETA  &  0.5\\
 GAMMA &  1.\\
 C     &  1.\\
  
 N     &   35\\
 NSMPL &   500000\\
 NVOID &   5000\\
  
 ZSTEP &  2.1 1/GeV \\ 
 SSTEP &  4.5 1/GeV$^2$\\
  
 EPSA  &  0.10E-03 GeV$^{-1}$ \\ 

 IDUM  &  1\\
 Initial coordinates: & 0  0  0  0 GeV$^{-1}$\\
                      & 1  1  0  0 GeV$^{-1}$\\
                      &-1  1  0  0 GeV$^{-1}$\\
 Final coordinates: & 0  0  0 35 GeV$^{-1}$\\
                    & 1  1  0 35 GeV$^{-1}$\\
                    &-1  1  0 35 GeV$^{-1}$\\
  
 INTOUT&  1\\
  
 z update \%&   50.33\\
  
 s update \%&   53.63\\
  
Bound state mass & 0.97$\pm$0.001 GeV\\
\hline
\end{tabular}

\end{document}